\documentclass{emulateapj}
\begin{document}
\title{Detection with RHESSI of high frequency X-ray oscillations in the tail of
  the 2004 hyperflare from SGR 1806-20} \author{Anna L. Watts$^{1,2}$  and Tod
  E. Strohmayer$^{1,3}$}  \altaffiltext{1}{Exploration of the Universe Division.,
NASA/GSFC, Greenbelt, MD 20771}
\altaffiltext{2}{Current address:  Max Planck Institut f\"ur
  Astrophysik, Karl-Schwarzschild-Str.1,
85748 Garching, Germany; anna@mpa-garching.mpg.de}
\altaffiltext{3}{Email: stroh@milkyway.gsfc.nasa.gov}
\begin{abstract}

The recent discovery of high frequency oscillations in
giant flares from SGR 1806-20 and SGR 1900+14 may be the first direct detection of  
vibrations in a neutron star crust. If this interpretation is correct it
offers a novel means of testing the neutron star equation of state, crustal
breaking strain, and magnetic field configuration.  Using timing data
from RHESSI, we have confirmed the detection of a 92.5 Hz Quasi-Periodic
Oscillation (QPO) in the tail of the SGR 1806-20 giant flare. We also
find another, stronger, QPO at higher energies, at 
626.5 Hz.  Both
QPOs are visible only at particular (but different) rotational phases,
implying an 
association with  a specific area of the neutron star surface or
magnetosphere.  At lower frequencies we confirm the detection of an 18
Hz QPO, at the same rotational phase as the 92.5 Hz QPO, and report
the additional presence of a broad 26 Hz QPO. We are however unable to
make a robust confirmation of the presence of a 30 Hz QPO,
despite higher countrates.  We
discuss our results in the light of neutron star 
vibration models.

\end{abstract}

\section{Introduction}

The Soft Gamma Repeaters (SGRs), objects that exhibit recurrent bouts of
gamma-ray flare activity,  are thought to be magnetars - neutron stars
with magnetic fields greater than $10^{14}$ G \citep{dun92, tho95,
  woo04}. On rare occasions SGRs exhibit giant flares, hugely
energetic events with peak fluxes in the range $10^{44} - 10^{46}
\mathrm{ergs~s}^{-1}$.    The giant flares consist of
a short spectrally hard initial peak, followed by a softer decaying tail lasting
a few hundred seconds.  Pulsations with periods of several seconds
are visible in the tail, and reveal the neutron star spin period.
Their presence is thought to be due to a fireball of ejected
plasma, trapped near the stellar surface by the strong magnetic field \citep{tho95}.

Three giant flares have been detected since the advent of 
satellite-borne high energy detectors: in 1979, from SGR 0526-66
\citep{maz79}; in
1998, from SGR 1900+14 \citep{hur99}; and in 2004 from SGR
1806-20 \citep{ter05, pal05}.  The catastrophic magnetic instability
that powers the giant flares is thought to be associated with large-scale
fracturing of the neutron star crust \citep{flo77, tho95,
  tho01, sch05}. This will
almost certainly excite global seismic vibrations \citep{dun98}:  terrestrial seismologists regularly observe such global modes
after large earthquakes such as the 2004 Sumatra-Andaman 
event \citep{par05}.  

The 2004 flare from SGR 1806-20 was the most energetic ever recorded.
Analysis of data from the {\it Rossi X-ray Timing Explorer} (RXTE)
by \citet{isr05} revealed a transient 92.5 Hz 
Quasi-Periodic Oscillation (QPO) in the tail of the
flare, associated with a particular rotational phase.  The presence of
18 and 30 Hz features was also 
suggested, although with lower significance. \citet{isr05} suggested
that the 30 Hz and 92.5 Hz QPOs might be toroidal shear vibrations
of the crust \citep{sch83, mcd88, str91, dun98, pir05}. These modes are
particularly promising in terms of both ease of excitation and
coupling to the magnetic field to modulate the X-ray lightcurve \citep{bla89}.

Motivated by this work, \citet{str05} re-analysed RXTE
data for the 1998 flare from SGR 1900+14.  QPOs were found at 28,
54, 84 and 155 Hz, again associated with a particular rotational
phase.  These QPOs can plausibly be identified with a series
of toroidal modes.  The association with a rotational phase away
from the main peak (where emission from the surface is obscured by the trapped
fireball) implies that we are seeing emission from a particular area
of the stellar surface or bundle of magnetic field
footpoints. 

The frequencies of crustal modes such as the toroidal modes depend on
neutron star mass and radius, crustal
rigidity, and magnetic field configuration \citep{dun98}. For this
reason the high frequency QPOs 
offer a powerful new
diagnostic of neutron star properties provided that an accurate mode
identification can be made (see \citet{str05} for an
initial attempt to constrain SGR parameters). 

In this paper we present a timing analysis of the SGR 1806-20
 hyperflare using data from the {\it Ramaty
 High Energy Solar Spectroscopic Imager} (RHESSI), and discuss our
results in the light of the crustal vibration model.   

\section{Observations and Data Analysis}

RHESSI is a solar-pointing satellite whose primary objective is the
study of solar flares in the energy range 3 keV - 17 MeV
\citep{lin02}.  The detection system comprises nine
cyrogenically-cooled Germanium detectors, divided into front and rear
segments.  For solar flares all direct photons under 100 keV should be
stopped in the front segments.   Most direct photons above 100 keV are
captured in the 
   higher volume rear segments. Lower energy photons reach the
rear detectors only indirectly, by scattering (there is a Beryllium scatterer
embedded in the front segments, intended for use in polarization
studies).  The rear segments also capture albedo
flux from the Earth. 

\begin{figure}
\begin{center}
\includegraphics[width=8.5cm, clip]{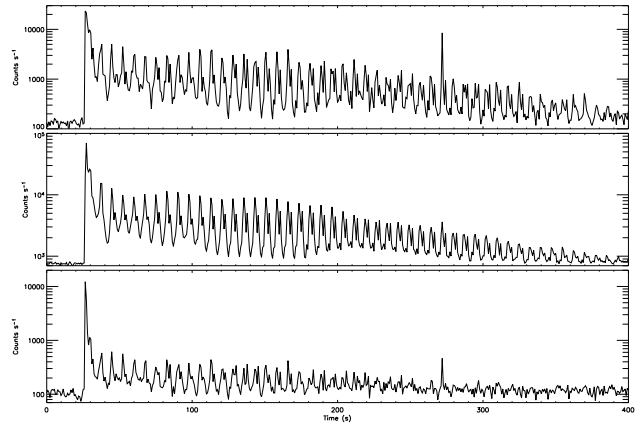}
\end{center}
\caption{Light curves.  Top:  Front segments, 25-100 keV band. Center:
Rear segments, 25-100 keV band.  Bottom:  Front segments, 100-200 keV
band. The plots show the main peak and decaying tail with the 7.6s
double-peaked pulse profile.  The spike in the front segments at
270s is due to the removal of an attenuator. Zero time
corresponds to 21:30 UTC on December 27 2004.} 
\label{f1}
\end{figure}

The SGR 1806-20 hyperflare was  $\approx 5^\circ$ off-axis for RHESSI
 \citep{bog04}.  The detector saturated during the peak,
 but recorded the decaying tail in its entirety
(\citet{hur05}; Figure \ref{f1}).  Although the flare was not directly
 in the RHESSI field of 
 view, most photons in the  
front segments would have been direct.  Given RHESSI's native time
 resolution of 1 binary $\mu$s ($2^{-20}$s) these events are clearly
 suitable for high frequency timing analysis.  The rear segment
 flux, by contrast, comprises scattered photons from the front
 segments, 
direct photons entering through the walls of the
spacecraft, and albedo flux.  The latter, which could be as much as
 40-50\% of the 
direct flux in the energy range of interest \citep{mcc04}, has a severe impact
on timing analysis.  At the time of the flare 
RHESSI was passing the limbs of the Earth (as viewed from the
SGR). Albedo flux is
limb-brightened, particularly if the incoming flux is polarized
\citep{wil05}.  This means that a large fraction of the detected photons
could have incurred additional delays of up to $\approx 0.02$ s,
smearing out signals above $\approx 50$ Hz.  Note that although
 countrates in the rear 
segments exceed those recorded by RXTE, countrates in the front
segments are slightly lower.  It should also be noted that scattering
 from the spacecraft walls and the Earth will cause the photon
 energies recorded by RHESSI, particularly in the rear segments, to
 deviate from the true energies of the incident photons.  Quantifying
 this effect precisely is extremely difficult. For this reason we use
 broad energy bands in our analysis, and urge some care in
 interpreting the recorded photon energies. 

We started by extracting event lists from the RHESSI data, excluding
only events occurring in a 2s period $\approx 270$s after the
peak of the flare when an attenuator is removed (the associated spike
introduces spurious variability, particularly in the front
segments). Timing analysis was carried out using the $Z^2_n$ statistic
\citep{buc83, str99}.  \citet{isr05} showed that the presence of the
high frequency signals was dependent on the phase of the 7.6s
rotational pulse; the signals appeared most strongly at phases away from the
main peak.  Similar phase-dependence was also observed in the SGR 1900+14 hyperflare \citep{str05}.  As such we have conducted a phase-dependent analysis.  

We searched for phase-dependent QPOs by folding data of a given
rotational phase from $N_p$ pulses, generating power spectra that
are averaged to a frequency resolution $\Delta \nu$.  The distribution of noise
powers is a $\chi^2$ distribution with 2$N$ degrees of freedom, 
where $N = N_p \Delta \nu P \Delta \Phi$, $P$ is the
rotational period and 
$\Delta \Phi$ is the phase window under consideration ($0 < \Delta
\Phi \le 1$).  We searched over a range of $\Delta \Phi$, $N_p$, and
energy bands for any signals with significance $> 3\sigma$.  

We started by searching for signals in the range 50-1000 Hz, using
only data from the front segments. In this range the noise profile is
Poissonian.  We find only two signals that meet our search criterion.

\begin{figure}
\begin{center}
\includegraphics[width=8.5cm, clip]{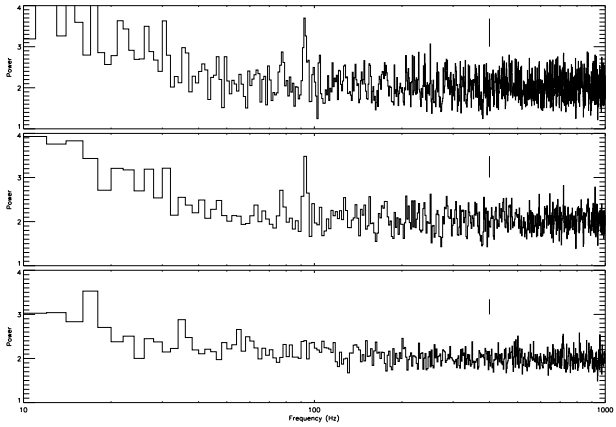}
\end{center}
\caption{Average power spectra from 2.27 s intervals (0.3 cycles)
centered on different rotational phases, computed using photons from
the front segments with recorded energies in the range 25-100 keV.
The upper curve was 
computed using 15 successive 2.27 s intervals, $\approx 150-260$ s
after the main flare, at a rotational phase that includes the secondary
peak and part of the DC phase. The frequency resolution
is 1 Hz.  The 
middle curve shows the same spectrum with 
2 Hz frequency resolution.  The QPO at 92.5 Hz is clearly visible.
The lower curve is for the same time
period but is an average of rotational phases $\pm$ 2.27 s away from
the 92.5 Hz signal phase:  no QPOs are detected.  Characteristic error
bars are shown for each spectrum.} 
\label{f2}
\end{figure}

The first, for photons with recorded energies in the range 25-100 keV, is the
QPO at 92.5 Hz previously reported by \citet{isr05}, 
shown in Figure \ref{f2}.  This signal, which we detect only at a
rotational phase away from the main peak, is strongest
$\approx 150-260$ s after the initial flare. As noted by
\citet{isr05}, this occurs in conjunction with an increase in unpulsed
emission.  At $\Delta \nu = $1 Hz the QPO is resolved; at $\Delta \nu = $2 Hz it is not.   We
estimate the significance of the $\Delta \nu = $2 Hz power using a
$\chi^2$ distribution  
with 68 degrees of freedom, which is the distribution expected based
on the number of independent frequency bins and pulses averaged.  The
peak at 93 Hz has a single trial probability of $2 \times 10^{-7}$.
Applying a correction for the number of frequency bins, independent
time periods, and rotational phases searched we arrive at a
significance of $\approx 1\times 10^{-3}$.  That this is lower than the
significance reported by \citet{isr05} is to be expected, given that
the RHESSI front segment countrate is lower than that 
of RXTE.  A search for the signal in the RHESSI rear segments indicates 
that the signal has indeed been smeared out due to albedo
flux.  Fitting the QPO with a Lorentzian profile we find a centroid
frequency 92.7 $\pm$ 0.1 Hz, with coherence value Q of 40.  The
integrated RMS fractional amplitude is 10 $\pm$ 0.3 \%, in good
agreement with \citet{isr05}.

The independent detection with RHESSI of the 92.5 Hz QPO is a strong
confirmation of the RXTE findings.  Using the significance
quoted by \citet{isr05}, we can compute the probability of getting
two apparent detections at the same frequency, time and phase, due to
noise alone, given the number of 
trials. If we do this we find that the detection of the 92.5 Hz 
QPO has a combined significance of $>6\sigma$, an extremely
robust result.  

\begin{figure}
\begin{center}
\includegraphics[width=8.5cm, clip]{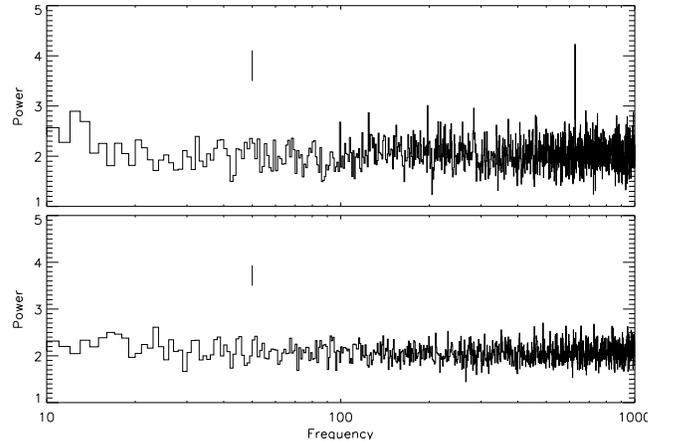}
\end{center}
\caption{Average power spectra from 2.27 s intervals (0.3 cycles)
centered on different rotational phases, computed using photons from
the front segments with recorded energies in the range 100-200 keV.
The upper curve was 
computed using 19 successive 2.27 s intervals, $\approx 50-200$ s
after the main flare, for a rotational phase centered on the main
peak. The frequency resolution is 1 Hz, and the QPO is 
clearly visible at 626.5 Hz.   The lower curve is for the same time
period but is an average of rotational phases $\pm$ 2.27 s away from
the main peak:  no QPOs are detected. Characteristic error
bars are shown for each spectrum.} 
\label{f3}
\end{figure}

\begin{figure}
\begin{center}
\includegraphics[width=8.5cm, clip]{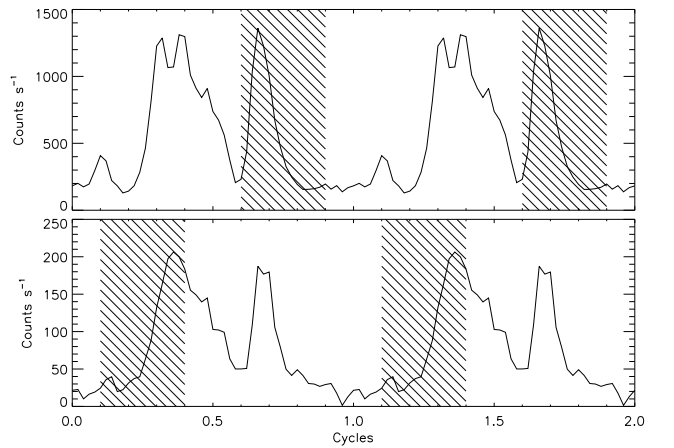}
\end{center}
\caption{Folded pulse profiles (background subtracted), showing
the rotational phase where the high frequency 
QPOs are detected.  The top panel shows the average pulse profile for
photons from the front segments, in
the 25-100 keV band, for
the period where the 92.5 Hz QPO is detected most strongly (2 full
cycles are shown). 
The lower panel shows the average pulse profile for
photons from the front segments, in the 100-200 keV
band, for the period when the 626.5 Hz QPO is detected most strongly.
The hatched regions indicate the rotational phases where the QPOs are
detected. }  
\label{f4}
\end{figure}

The second detection, for photons with recorded energies in the range
100-200 keV, is of a narrower QPO at 626.5 Hz (Figure \ref{f3}). This
signal is strongest in the period 50-200 s after the main flare.
As before, the signal is only detected at certain
rotational phases; but in this case the rotational phase where the
signal is strongest is centered on the leading edge of the main peak
(Figure \ref{f4}). We
estimate the significance of the peak using a $\chi^2$ distribution 
with 114 degrees of freedom, which is the distribution expected based
on the number of independent frequency bins and pulses averaged.  The
peak at 626.5 Hz has a single trial probability of $7.7 \times
10^{-9}$. Applying a correction for the number of frequency bins, independent
time periods, and rotational phases searched we arrive at a
probability of chance occurrence of $\approx 6.6 \times
10^{-5}$. Fitting the QPO with a Lorentzian profile we find a centroid 
frequency 626.46 $\pm$ 0.02 Hz, with coherence value Q of 790.  The
integrated RMS fractional amplitude is high, at 20 $\pm$ 3 \%.

\begin{figure}
\begin{center}
\includegraphics[width=8.5cm, clip]{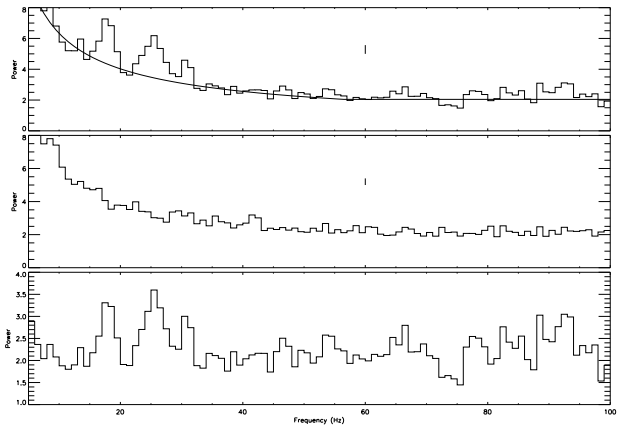}
\end{center}
\caption{Average power spectra from 2.27 s intervals (0.3 cycles)
centered on different rotational phases, computed using photons from
the front and rear segments with recorded energies in the range 25-100 keV.
The upper curve was 
computed using 22 successive 2.27 s intervals, $\approx 60-230$ s
after the main flare, for the rotational phase where the 92.5 Hz QPO is
observed. Several possible QPOs are apparent at low
frequencies, and we show the best-fit power law plus constant model of
the continuum.  The center curve is for the same time
period, but is an average of rotational phases $\pm$ 2.27 s away from
the main peak:  no QPOs are detected.  In the lower panel we have
renormalised the power spectrum using the continuum model. Broad QPOs
at 18 Hz and 26 Hz are readily apparent, as 
is a weaker feature at 30 Hz. Note that there is no longer a distinct
QPO peak at 92.5 Hz, as there was in Figure 2, although there is a
broader feature at this frequency.  The fact that the peak at 92.5 Hz
has been smeared out illustrates the effect of albedo flux in the rear
segments on high frequency signals. The frequency resolution in all panels
is 1 Hz. Characteristic error
bars are shown.  }
 \label{f5}
\end{figure}

Extending our search to lower frequencies, in the range 1-50 Hz, we
can now use data from both the front and rear segments.  In this
range the noise spectrum is not Poissonian.  If we focus only on the
time period 200-300s 
after the main flare, where \citet{isr05} reported possible QPOs at 18 and 30
Hz, we do not find features that meet our search criterion.
Extending our search to earlier times, 
however, we do find evidence for low frequency features at the
rotational phase where the 92.5 Hz QPO was found (Figure \ref{f5}).
At 1 Hz resolution we find broad features at $\approx$ 18 and 26 Hz, with a
slightly weaker feature at $\approx$ 30 Hz. Assigning a significance
to these putative detections requires us to fit the noise powers.  We
fit a constant plus power law model, excluding the candidate peaks 
to avoid biasing the fit.  We then divide by the continuum model and
multiply by 2 to
obtain the normalised spectrum shown in the lower panel of Figure
\ref{f5}.  We compute significances using a 2 Hz resolution power
spectrum (not shown), to take into account the breadth of the candidate
QPOs. If we neglect the uncertainty associated with the normalisation,
and assume that the noise is Poisson, the single trial probabilities of the powers
measured at 18 Hz and 26 Hz are $3.9\times 10^{-8}$ and
$2.7\times 10^{-9}$ respectively.  Accounting for the number of
trials, the probabilities of chance occurrence are $8.8\times 10^{-6}$
for the 18 Hz QPO and $6.1\times 10^{-7}$ for the 26 Hz QPO.  The
uncertainty associated with the noise model will reduce the
quoted significances by some factor, but unless the noise model is seriously in
error, these detections are robust (see \citet{van89} and
\citet{isr96} for further discussion of this issue).  An 18 Hz feature 
was seen by \citet{isr05}; it would be
interesting to see if a closer analysis of the RXTE dataset reveals
the 26 Hz feature. Fitting a Lorentzian profile we find for the lower
frequency QPO a centroid frequency 17.9 $\pm$ 0.1 Hz, Q of 10, and
integrated RMS fractional amplitude 4.0 $\pm$ 0.3 \%.  For the higher
frequency QPO we find a centroid frequency 25.7 $\pm$ 0.1 Hz, Q of 9
and an integrated RMS fractional amplitude of 5.0 $\pm$ 0.3 \%.  

 The weaker 30 Hz
feature does not meet the $3\sigma$ criterion at either
1 Hz or 2 Hz resolution.   As such we are
unable to make a robust confirmation of the 30 Hz QPO suggested by 
\citet{isr05}, despite higher countrates.  One might boost the
significance of this feature by combining RXTE and RHESSI datasets;
but since the time window reported by \citet{isr05} is different, some
caution is warranted.

\section{Discussion}

The strong rotational-phase dependence of the QPOs in the tail of the 
SGR 1806-20 hyperflare provides compelling
evidence that all of the oscillations are assocated with
particular regions of the stellar surface or magnetosphere.  In this
section we discuss the candidate mechanisms in more detail.

\citet{isr05} identified the 92.5 Hz QPO with the $l=7, n=0$ toroidal mode of the
crust.  This intepretation was supported by the
apparent detection of a QPO at 30 Hz, the expected frequency of the 
fundamental $l=2$ mode.  Although we cannot confirm the detection
of the 30 Hz QPO, the interpretation of
the 92.5 Hz QPO is probably
secure. If the magnetic field of $\sim 10^{15}$ G inferred from
timing analysis \citep{woo04} is accurate, then for the 92.5 Hz QPO to be an $l=6$
toroidal mode the stellar mass would have to be $< 1M_\odot$ for
even the softest equation of state.  To be an $l=8$ mode, the
neutron star mass would have to exceed that of SGR
1900+14 by
$>0.3 M_\odot$ (see the discussion in \citet{str05}). An alternative to the toroidal modes, with a frequency in the correct
range, is the 
crustal interface mode, but its behavior
in the presence of strong magnetic fields requires further study
\citep{pir05a, pir05}.

The detection of a QPO at 626.5 Hz is exciting, as there are several
crustal modes that could have frequencies in this 
range.  The most likely candidate is the $n=1$ toroidal mode;
the 
detected frequency agrees extremely well with the most recent models
\citep{pir05}.  This would be fascinating in terms of energetics; the
energy required to excite $n>0$ modes is orders of 
magnitude larger than that required to excite an $n=0$ mode, a
testament to the strength of this particular 
flare.  Other candidates include the crustal spheroidal modes
or the crust/core interface modes \citep{mcd88}, although the effect of
a strong magnetic field on these modes remains to be calculated. If
this mode is indeed a higher radial overtone, it could allow us to
determine the depth at which the fracture occurred in the crust.  

The nature of the 18 and 26 Hz QPOs is less clear.  The frequencies
are too low for the fundamental crustal toroidal mode unless the neutron star
parameters are very extreme.  A torsional mode of the core, restored by the poloidal magnetic field
\citep{tho01}, was suggested by \citet{isr05} as a candidate mechanism
for the 18 Hz QPO. Whether such modes could explain both of the QPOs is a matter for further study.  

The fact that the various QPOs are strongest at different times,
suggests that we may be seeing two 
separate fracture/reconnection events.  The first, associated with the
main flare, excites the 626.5 Hz QPO.  The
second, associated with the late time boost in unpulsed emission,
excites the 92.5 Hz QPO.   The idea that we might be seeing
separate fracture sites is given additional weight by the observation that
the two QPOs are 
strongest at different rotational phases.  Such sequential rupturing is often
observed in terrestrial seismology. It is interesting that the
18 and 26 Hz QPOs are strongest at the rotational phase where the 92.5
Hz QPO subsequently appears.  One might speculate that the magnetospheric
oscillations associated with the low frequency QPOs (which are
triggered by the main flare) slowly weaken the crust at this point
and trigger the second fracture.  

Understanding the means by which the modes couple to the magnetic 
field, and hence 
modulate the X-ray lightcurve, is now critical.  
The fact that the modes appear in different energy bands is
intriguing, as is the fact that the 626.5 Hz QPO appears at a
rotational phase when a large part of the surface should be
obscured by the trapped fireball. Much
theoretical effort is clearly still required, but prospects
for neutron star asteroseismology are bright.

\acknowledgements
We would like to thank Brian Dennis, Kim Tolbert, David Smith and
in particular Richard Schwartz for advice on RHESSI data analysis. ALW acknowledges
support from a National Research Council Resident 
Research Associateship.

\end{document}